\documentclass[twocolumn]{jpsj3}
\usepackage{txfonts}
\usepackage{bm}% bold math
\usepackage{color}
\usepackage{ulem}
\usepackage[dvipdfmx]{graphicx}% Include figure files
\usepackage{dcolumn}% Align table columns on decimal point
\usepackage{mathrsfs}
\usepackage{ulem}
\usepackage{xcolor}
\DeclareRobustCommand{\erase}{\bgroup\markoverwith{\textcolor{red}{\rule[.5ex]{2pt}{0.4pt}}}\ULon}

\title{$^{139}$La-NMR study of spin-dynamics coupled with hole mobility in $T$*-type La$_{0.86}$Eu$_{0.86}$Sr$_{0.28}$CuO$_{4-\delta}$}

\author{Takanori  Taniguchi\thanks{Electronic address: taka.taniguchi@imr.tohoku.ac.jp, Present address: Institute for Materials Research, Tohoku University, Sendai 980-8577, Japan}$^{1,2}$, Shunsaku Kitagawa$^{1}$, Kenji Ishida$^{1}$, Shun Asano$^{2,3}$, Kota Kudo$^{2,3}$, Motofumi Takahama$^{2,3}$, Peiao Xie$^{2,3}$, Takashi Noji$^{4}$, and Masaki Fujita$^{2}$}
\inst{%
$^{1}$Department of Physics, Kyoto University, Kyoto 606-8502, Japan.\\
$^{2}$Institute for Materials Research, Tohoku University, Sendai 980-8577, Japan.\\
$^{3}$Department of Physics, Tohoku University, Sendai 980-8578, Japan.\\
$^{4}$Department of Applied Physics, Graduate School of Engineering, Tohoku University, Sendai 980-8579, Japan.\\
}%

\date{\today}% It is always \today, today,
\abst{
In $T$*-type cuprate oxides with five oxygen coordination, little is known about the relationship between the spin correlations and dope carriers. 
We performed $^{139}$La-nuclear magnetic resonance (NMR) and electrical resistivity measurements on an as-sintered (AS) and oxidation annealed (OA) polycrystalline $T$*-type La$_{0.86}$Eu$_{0.86}$S$_{0.28}$CuO$_4$ (LESCO) to investigate its magnetic and superconducting (SC) properties. 
Upon cooling, the NMR spectrum of AS LESCO broadened below 3 K, at which the nuclear spin-lattice relaxation rate $1/T_1$ against the temperature exhibited a maximum, thereby indicating the appearance of static magnetism. The temperature dependence of $1/T_1$ between 3 K and 20 K was similar to that of the resistivity displaying the semiconducting behavior. 
Furthermore, the energy scale of the transport gap and spin-dynamics estimated was found to be comparable.  
These results suggest a close connection between the mobility of the doped carriers and low-energy spin-dynamics, as reported for lightly doped $T$-type La$_{2-x}$Sr$_x$CuO$_4$. 
In the OA SC sample, we confirmed the absence of a magnetic order and the Korringa relation above 10 K. Therefore, in the $T$*-type LESCO with $x$ = 0.28, the magnetic state coupled with holes drastically turns to the weakly correlated metallic state by oxidation annealing. 
}

\begin{document}
\maketitle
\section{\label{sec:level2}Introduction}
The relationship between magnetism and superconductivity in cuprate oxides has been one of the most fascinating topic of interest in condensed matter physics. 
Among cuprate superconductors, the 214 family exhibits a simple crystal structure and its chemical formula is described as $R_2$CuO$_4$ ($R$ = rare-earth element). 
As shown in Fig. 1, there are three types of structural isomers of the 214 family: the K$_2$NiF$_4$ structure with six oxygen coordination ($T$-type), the Nd$_2$CuO$_4$ structure with four oxygen coordination ($T^{\prime}$-type), and an intermediate structure between these two structures with a CuO$_5$ pyramidal having five oxygen coordination ($T$*-type) around the Cu ions\cite{Uchida_1987, Takagi_1987, Tokura_1989, Akimitsu_1988}. 
The presence and position of apical oxygen vary the energy of the charge transfer gap~\cite{Arima1991, Tshukada2006} and hopping parameters~\cite{Ikeda2009}. Thus, the spin and charge correlations could be related to the coordination of Cu. 

Among the isomers of $R_2$CuO$_4$, the physical properties of the hole-doped $T$-type La$_{2-\rm x}$(Sr, Ba)$_{\rm x}$CuO$_4$ have been extensively studied because of the controllable hole density ($p$) via elemental substitution at the $R$ sites and the availability of single crystals. 
In general, superconductivity emerges with varying spin correlations by doping a sufficient number of holes into the parent Mott insulator La$_2$CuO$_4$\cite{Yamada_1998, Fujita_2012, Weidinger_1989, Julien_2003, Curro_2000, Hunt_2001}. The spin correlations are related to the doped holes in a wide $x$ range depending on the value of $p$; low-energy spin-dynamics is coupled with hole mobility in the lightly doped (LD) region\cite{Ishida_2004, Hucker_2004}, and the spatially modulated spin density appears with a one-dimensional alignment of the doped holes in the superconducting (SC) phase~\cite{Tranquada_1995, Yamada_1998}. 
Superconductivity and the intertwining of the spin and charge (hole) orders~\cite{Wen2019, Singer2020} have garnered considerable attention as a new form of pairing, i.e., the pair density wave~\cite{Fradkin_2015, Agterberg_2020, Huang_2021}. 

%Fig.1%%%%%%%%%%%%%%%%%%%%%%%%%%%%%
\begin{figure}[b]
\begin{center}
\includegraphics[width=8cm,clip]{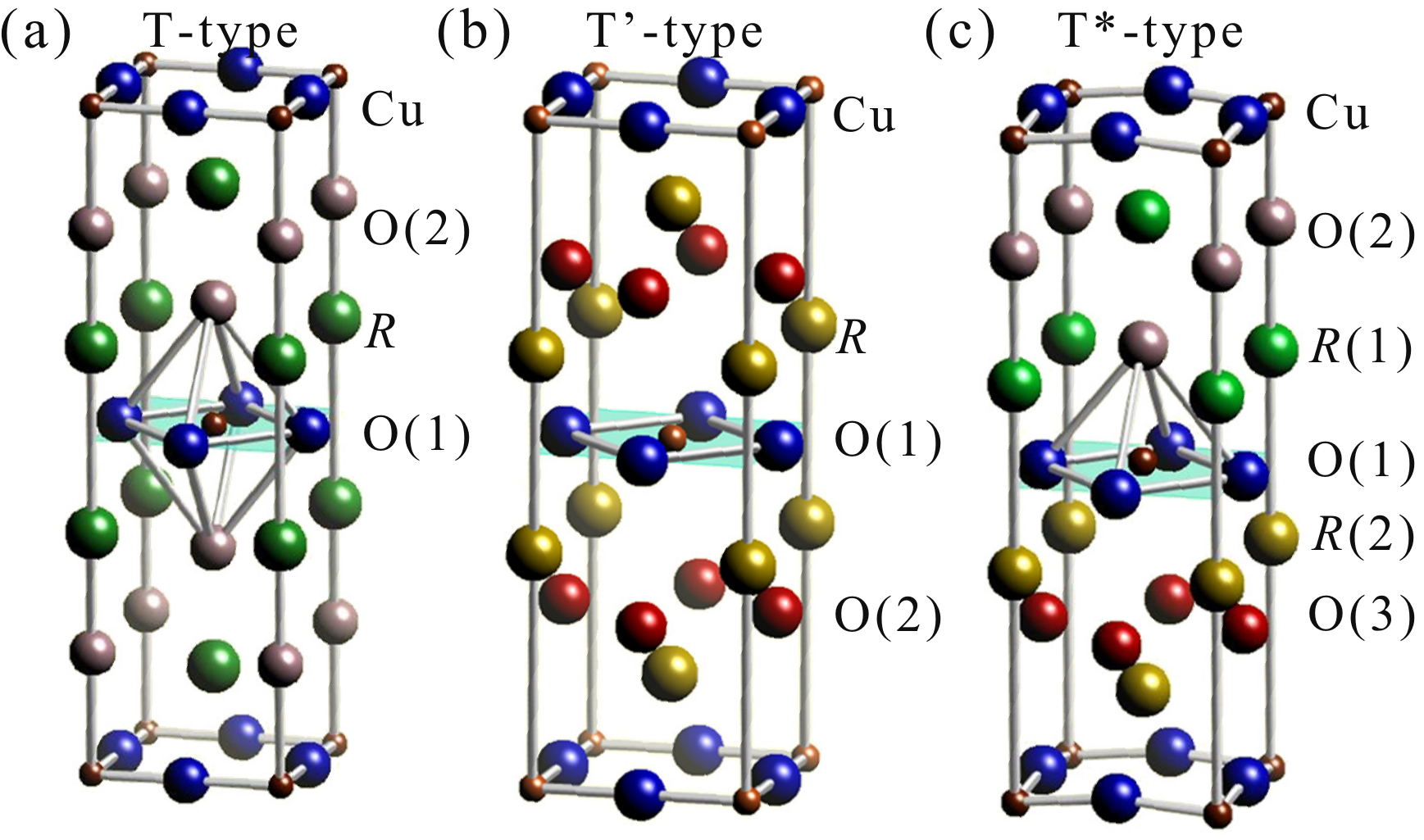}% Here is how to import EPS art
\caption{\label{fig:mag_para} (Color online) Crystal structure of (a) $T$-type and (b) $T$$^{\prime}$-type cuprates with $R$ denoting rare-earth sites. (c) Crystal structure of $T$*-type La$_{1-x/2}$Eu$_{1-x/2}$Sr$_{x}$CuO$_4$ ($R(1)=\rm{La, Sr, Eu}$ and $R(2)=\rm{La, Eu}$). Semitransparent sheet represents a CuO$_2$ plane. 
}
\end{center}
\end{figure}\rm

Compared to the $T$-type LSCO, the physical properties of the hole-doped $T$*-type cuprate are still unclear. 
The characteristic in the crystal structure of the $T$*-type cuprate is that the atomic arrangement in one half of the unit cell is identical to that of the $T$-type structure, and the other half exhibits a $T$$^{\prime}$-type structure (Fig. 1(c)). Therefore, two rare-earth sites are present in the $T$*-type cuprate, occupying the same Wyckoff positions (2c) with point symmetry $4mm$ with the P4/nmm (No. 129) space group\cite{Akimitsu_1989, Sawa_1988}. 
The oxidation annealing under high pressure is necessary for the emergence of superconductivity in the $T$*-type cuprate\cite{Akimitsu_1988}. %
In this context, a structural study reported that the partially existing oxygen vacancies at the apical site (O(2) in Fig. 1(c)) in an as-sintered (AS) compound can be repaired through oxidation annealing. 
Superconductivity is considered to appear by eliminating defects. 
Therefore, the relationship between spin and charge correlations can be investigated via the annealing effect, which is similar to the electron-doped $T$$^{\prime}$-type cuprate. 

%Fig.test%%%%%%%%%%%%%%%%%%%%%%%%%%%%%
\begin{figure}
\begin{center}
\includegraphics[width=7.0cm,clip]{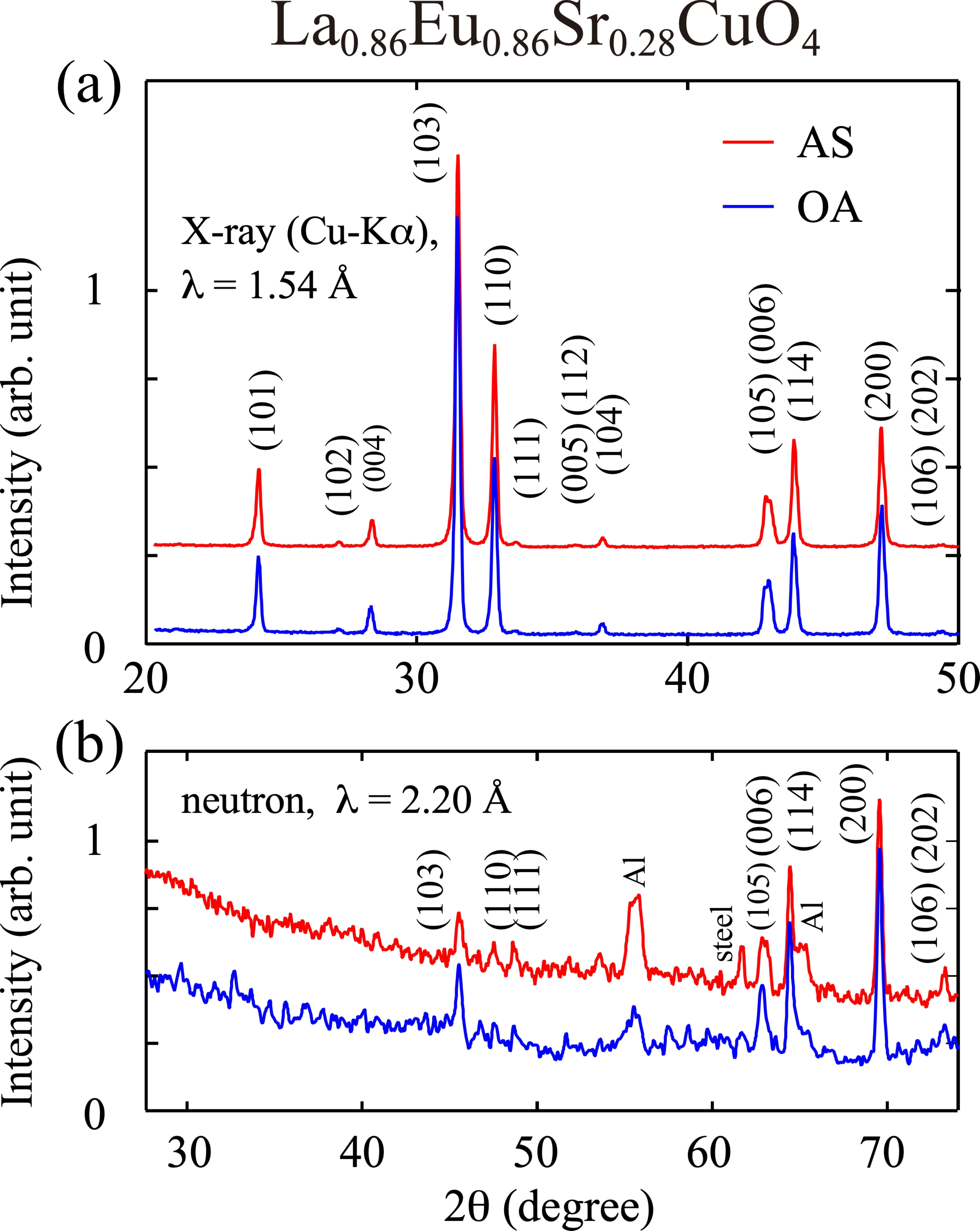}% Here is how to import EPS art
\caption{\label{fig:mag_para} (Color online) (a) X-ray and (b) neutron diffraction patterns of AS and OA La$_{0.86}$Eu$_{0.86}$Sr$_{0.28}$CuO$_4$. Al and steel in Fig. 2(b) represent the background from aluminum and steel around the sample. }
\end{center}
\end{figure}\rm 

Recently, we investigated the magnetic and SC properties of powder samples of $T$*-type La$_{1-\rm x/2}$Eu$_{1-\rm x/2}$Sr$_{\rm x}$CuO$_4$ (LESCO) with $0.14\leq x\leq0.28$\cite{Fujita_2018, Asano_2019}. Asano {\it et al}. clarified the semiconducting behavior of electrical resistivity and the spin-glass (SG) anomaly based on the hysteresis behavior in magnetic susceptibility for AS LESCO\cite{Asano_2019}. Furthermore, muon spin rotation/relaxation ($\mu$SR) measurements revealed the emergence of a bulk magnetic order in the wide $x$-region of AS LESCO below $\sim$7 K\cite{Asano_2019}. 
Oxidation annealing induces superconductivity in all samples by drastically reducing the magnitude of resistivity below room temperature. 
Interestingly, $T_{\rm c}$ for the oxidation annealed (OA) LESCO with $0.14\leq x\leq0.28$ continuously increased with the decreasing $x$, and both the magnetic susceptibility and $\mu$SR measurements did not evidently detect any magnetic order in the OA samples. As the evaluated value of $p$ per Cu atom of OA LESCO from oxygen $K$-edge X-ray absorption spectroscopy (XAS) measurement was lower than the value of $x$~\cite{Asano_2020}, $T_{\rm c}$ apparently increased in the LD region toward zero doping, unlike the dome-shaped SC phase of LSCO. The emergence of superconductivity in the parent $R_2$CuO$_4$ has been reported for $T$$^{\prime}$-type cuprates~\cite{Matsumoto_2009, Takamatsu_2012, Sunohara_2020}, and the mechanism of superconductivity without elemental substitution attracts much attention. %, although the comprehensive study on physical properties is still limited. 
Thus, the relationship between spin and charge dynamics in $T$*-type cuprates, which potentially exhibits superconductivity near zero-doping, should be clarified for understanding the impact of oxygen coordination on the electronic state.

In this paper, the results of the $^{139}$La-nuclear magnetic resonance (NMR) and the electrical resistivity ($\rho$) measurements are presented on the AS and OA LESCO with $x$ = 0.28. The OA LESCO with $x$ = 0.28 was reportedly located in vicinity of the end point of the SC phase, whereas the AS sample exhibited SG-like magnetism at low temperatures\cite{Asano_2019}. Thus, the physical properties showing the significant variation can be investigated in LESCO with $x$ = 0.28 via the oxidation annealing. 

\section{\label{sec:level2}Sample Preparation and Experimental Details}
The AS samples of LESCO were synthesized using a solid-state reaction method as described in Refs. \citen{Fujita_2018} and \citen{Asano_2019}.
The OA samples were obtained by annealing the AS samples in high-pressure oxygen gas at 40 MPa under 500 $^\circ$C for 80 h. The increased oxygen content $\delta$ in the unit formula was estimated as $\sim$0.02 from the weight gain due to the annealing.

%Fig.3%%%%%%%%%%%%%%%%%%%%%%%%%%%%%
\begin{figure}
\begin{center}
\includegraphics[width=7.8cm,clip]{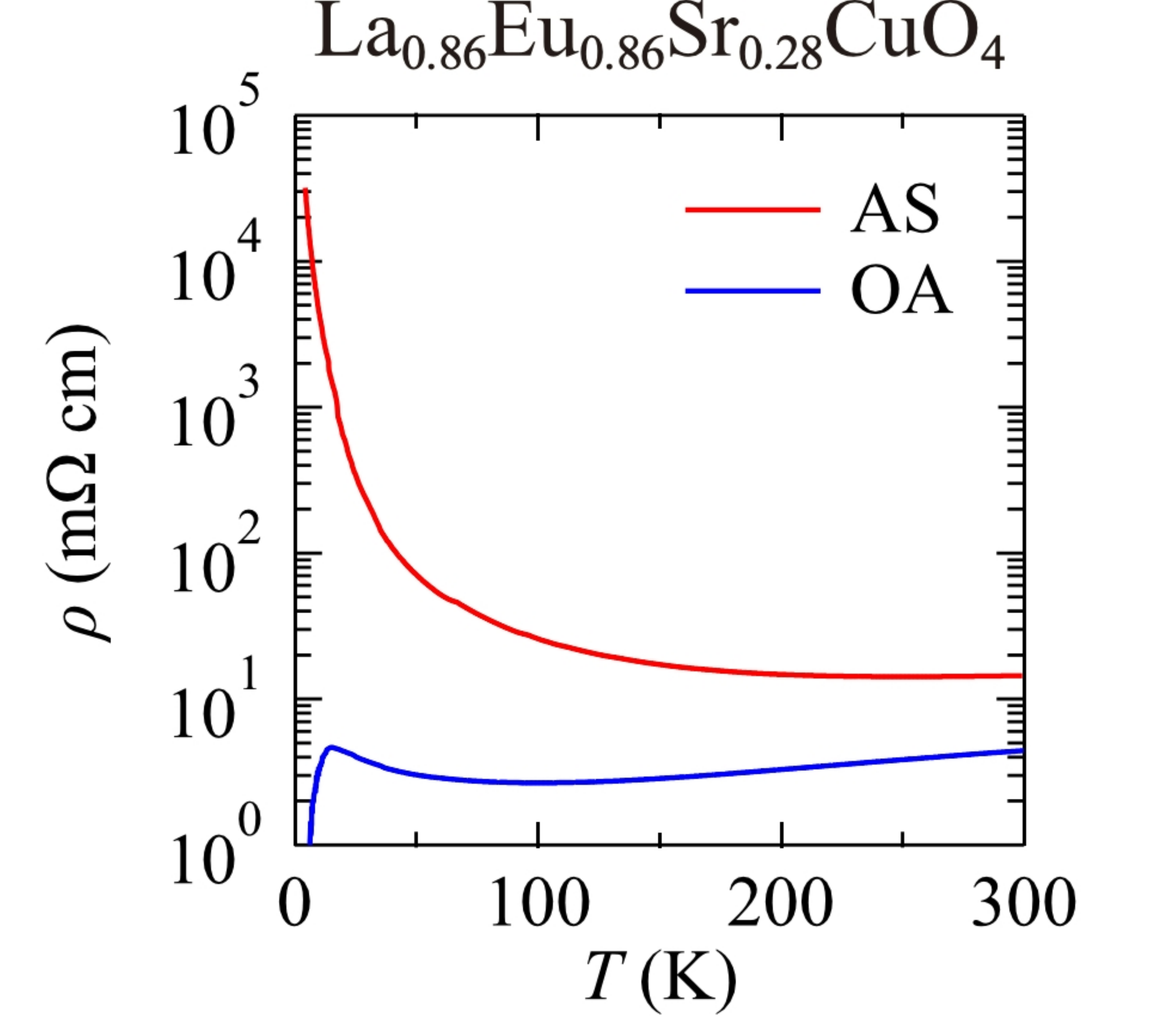}% Here is how to import EPS art
\caption{\label{fig:mag_para} (Color online) (a) Temperature dependence of $\rho$ in AS and OA La$_{0.86}$Eu$_{0.86}$Sr$_{0.28}$CuO$_4$. %(b) Temperature dependence of $\chi _{\rm ac}$ under various magnetic fields in OA La$_{0.86}$Eu$_{0.86}$Sr$_{0.28}$CuO$_4$. $T_{\rm c}$ is determined by linear fitting of diamagnetic signals at high and low temperatures, as denoted with black lines at the field of 1 T. The resultant $T_{\rm c}$ is indicated by arrow.  
}
\end{center}
\end{figure}\rm 

%Fig.4%%%%%%%%%%%%%%%%%%%%%%%%%%%%%
\begin{figure*}[t]
\begin{center}
\includegraphics[width=18cm,clip]{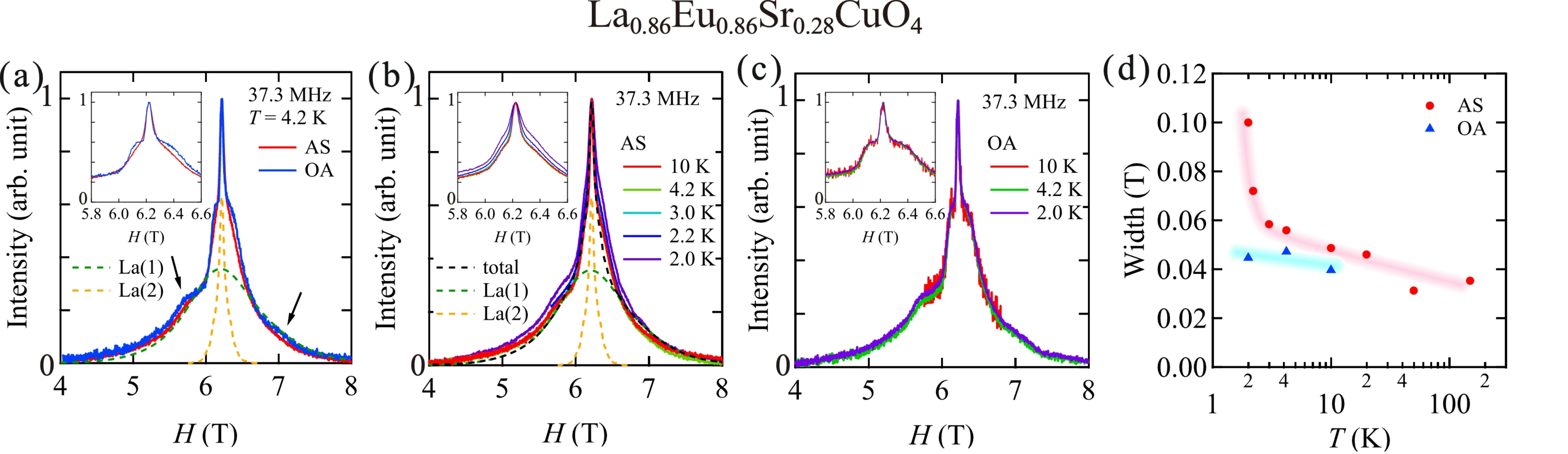}% Here is how to import EPS art
\caption{\label{fig:T1} (Color online) (a) $^{139}$La-NMR spectra in AS and OA La$_{0.86}$Eu$_{0.86}$Sr$_{0.28}$CuO$_4$ (LESCO) measured at 4.2 K and 37.3 MHz. The dotted lines denote simulated spectra convoluted with two Gaussian functions with an full-width of $\sim0.8$ and 0.06 T. Black arrows indicate magnetic fields where NMR spectra for two samples show the intensity difference. Temperature dependence of $^{139}$La-NMR spectra in (b)AS and (c)OA LESCO below 10 K at 37.3 MHz. Inset of each figure represents a portion of NMR spectra in field range between 5.8 and 6.6 T. (d) Temperature dependence of full-width at 4/5-maximum in the NMR spectra for AS and OA LESCO. Solid lines provide a guide to the eye. }
\end{center}
\end{figure*}
%\section{\label{sec:level3}Results}

The phase purity of the AS and OA samples was examined using X-ray and neutron powder diffraction measurements. 
In the X-ray diffraction measurements, Cu--K$\alpha$ radiation with the wavelength ($\lambda$) of 1.54 $ \AA$ was used to collect the diffraction profiles. In addition, we conducted neutron diffraction measurements using Ge(331) reflection ($\lambda = 2.20 $ \AA) on HERMES diffractometer at Japan Research Reactor No. 3 (JRR-3), Japan. As seen in Fig. 2(a), all the peaks in the X-ray diffraction pattern were assigned using the $T$*-type structure with $P4/nmm$ symmetry, and no evidence of the impurity phase was detected. 
The in-plane and out-of-plane lattice constants for AS sample were evaluated as 3.852(5) \AA~and 12.591(4) \AA, and those for the OA sample were 3.852(0) \AA~and 12.601(1) \AA, respectively. Thus, the $a$-axis reduced in size and the $c$-axis elongated owing to the oxidation annealing. Consistent results were obtained by neutron diffraction measurements (Fig. 2(b)). 

Moreover, we measured $\rho$ with a standard ac-four-probe method above 4.2 K. The NMR measurements were conducted after the zero-field-cooling process using $^4$He cryostats and 15 T SC magnet. 
%\textcolor{blue}{The pickup coil used in NMR and $\chi_{ac}$ measurements was identical. }
%
%
%
$^{139}$La-NMR frequency was maintained constant at 37.3 MHz, which represents the same frequency as the previous $^{139}$La-NMR measurement for the $T$$^{\prime}$-type La$_{1.8}$Eu$_{0.2}$CuO$_4$ (LECO) performed by Fukazawa {\it et al}\cite{Fukazawa_2017}. 
The $^{139}$La-NMR spectra were obtained by adding the Fourier transform spectra from the spin-echo signals that were measured with equally separated magnetic fields at a constant frequency.

\section{\label{sec:level2}Results and Discussion}
\subsection{\label{sec:level3}General behavior of resistivity}
The temperature dependence of $\rho$ for AS and OA LESCO with $x = 0.28$ is presented in Fig. 3(a), wherein the $\rho$ in the AS LESCO increased with the decreasing temperature, thereby indicating the semiconducting behavior. The magnitude of $\rho$ in the measured temperature range is reduced due to the annealing, and the evidence of superconductivity was observed with the transition temperature ($T_{\rm c}$) of $\sim$7 K.  
These results were consistent with the recent report \cite{Asano_2019}. %In Fig. 3(b), the temperature dependence of $\chi _{\rm ac}$ measured at various magnetic fields is shown for the OA sample. In particular, the $\chi _{\rm ac}$ decreased below $T_{\rm c} \sim 8$ K to signify the appearance of the SC state; the transition temperature was insensitive to the magnetic field up to 10 T. 

\subsection{\label{sec:level3}NMR spectral analysis}
The $^{139}$La-NMR spectra of the AS and OA samples at 4.2 K are illustrated in Fig. 4(a). 
The spectra for both the samples contained a sharp peak and an extreme broad peak ranging over several tesla. 
As mentioned earlier, two La sites in the $T$*-type LESCO correspond to the rare-earth ion sites in the $T$- and $T$$^{\prime}$-type cuprates. 
To gain insight into the microscopic information at La sites, we analyzed the spectra assuming two components. As resonance frequency ($\nu _Q$) is adequately sensitive toward the gradient of the electrostatic potential associated with the ions in the rare-earth sites, the difference in the crystal structure yields the individual values of $\nu _Q$. 
The NMR resonance frequency with 4-fold symmetry can be typically evaluated by determining the eigenvalues for the following Hamiltonian: 
\begin{equation}
\mathscr{H}=\sum_{i}\left\{-\gamma\hbar\boldsymbol{I}\cdot\boldsymbol{H^{i}}+\frac{1}{6}h\nu^{i}_{Q}\left(3I_{z}^{2}-\boldsymbol{I}^{2}\right)\right\}~~(i = T,~T^{\prime}),
\end{equation}
where the first term represents the Zeeman interaction with $h$ as the Plank constant, $\boldsymbol{I}$ as the nuclear spin, and $\gamma$ as the gyromagnetic ratio (6.0142~MHz/T for $^{139}$La). We analyzed the spectra illustrated in Fig. 4(a) by diagonalizing the Hamiltonian of Eq. (1) with the free parameters of electric field gradient at the La site $\boldsymbol{H^{i}}$ and resonance frequency $\nu^{i}_{Q}$ ($i = T,~T^{\prime}$). The $\nu^{i}_{Q}$ for the broad and sharp NMR spectra was obtained as $\sim$ 5 MHz and $\sim 0.8$ MHz, respectively. 
%
%, which were comparable to the corresponding values for the $T$-type LSCO and $T$$^{\prime}$-type LECO. 
%
%
The previous La-nuclear quadrupole resonance (NQR) and La-NMR measurements reported $\nu^{T}_{Q}$ of $\sim 6$ MHz for $T$-type LSCO ($0\leq x\leq0.018$)\cite{Ishida_2004, KitaokaJPSJ1987} and $\nu^{T'}_{Q}$ of $\sim 0.5$ MHz for $T$$^{\prime}$-type LECO\cite{Fukazawa_2017}. 
Therefore, based on the comparison with $\nu^{T}_{Q}$ and $\nu^{T'}_{Q}$, the broad (sharp) component in the $T$*-type LESCO could be attributed to the signal from La(1) (La(2)) site.

Next, we investigated the annealing effect on the NMR spectra. 
The spectra presented in Fig. 4(a) shows the intensity difference between the AS and OA samples around the magnetic fields marked with arrows. The difference can be more clearly seen in the inset, in which the intensity is shown in field range between 5.8 and 6.6 T. 
Upon comparing the spectra of the paramagnetic phase, the major deviations are predominantly associated with the structural variations caused by the annealing but not with the magnetism. 
Notably, no evident variation was observed in the NMR signals proximate to the peak position ($\sim$6.2 T). The peak position for either the broad component for La(1) site or the sharp component for La(2) site reflects $\left|I_z=-1/2\right\rangle\leftrightarrow\left|I_z=1/2\right\rangle$ transition, which is insensitive toward the electric field gradient. Thus, the enhancement of the intensity at particular fields suggest the variations in $\nu _Q$ due to the annealing. The spectral change observed on the broad component implies that the vacant apical oxygens around the La(1) site were repaired by the oxidation annealing. 

We furthermore examined the temperature dependence of the La-NMR spectra for the AS and OA samples. 
The NMR spectrum of the AS sample measured below 10 K are depicted in Fig. 4(b). The peak width varies with the temperature, whereas the peak field is insensitive toward the temperature, indicating that the Knight-shift is temperature independent below 10 K. 
To investigate the magnetism of the sample with avoiding structural effects on the spectrum, we focused the signal in the vicinity of sharp peak at $\sim$6.2 T. 
We evaluated the peak width from the field positions at which the intensity is 4/5 of the maximum in the raw data, and plotted the temperature dependence of the full-width in Fig. 4(d).
Upon cooling, the width gradually increases below 20 K owing to thermal contraction and/or weak Curie contribution, and it abruptly increases below 3 K, thereby reflecting the appearance of the static magnetism~\cite{Asano_2019}. 
In contrast, the NMR spectra of the OA sample are insensitive to temperatures below 10 K (Fig. 4(c)), which indicates the absence of static magnetism. Thus, the magnetic order disappears owing to oxidation annealing.

\subsection{\label{sec:level3}Temperature dependence of nuclear spin-lattice relaxation}

We evaluated the $1/T_1$ of $^{139}$La sites by measuring the recovery of the nuclear magnetization after a saturation pulse sequence. 
Although the fitting to the recovery data was challenging with the conventional theoretical equation\cite{Suter_1998}, the $1/T_1$ was phenomenologically estimated using the following stretch exponential equation: 
\begin{equation}
M\left(t\right)=M\left(\infty\right)-M_{0}\exp\left\{- \left(\frac{t}{T_{1}}\right)^{\beta}\right\}, \label{eq:T1}
\end{equation}
where the exponent value of $\beta$ is a quantity related to the inhomogeneity of the relaxation component and $M$ denotes the nuclear magnetization. 
The temperature dependence of $1/T_1$ and $\beta$ for the AS and OA samples are presented in Fig. 5(a) and 5(b), respectively. 
The $1/T_1$ of the AS LESCO displays a peak at 3 K, below which the $^{139}$La-NMR spectrum broadens owing to the appearance of the magnetic order. 
The $\beta$ reduces as the temperature dropped below 20 K, indicating an increase in the distribution of hyperfine fields at La sites. 
Similar inhomogeneous distribution of the magnetic field at the muon-stopping sites was reported based on the time spectra of the muon depolarization for the same AS sample\cite{Asano_2019}. 
Overall, the presence of such inhomogeneous magnetism is consistent with the appearance of the SG state. 
On the contrary, the OA sample displays no peak structure in the temperature dependence of $1/T_1$, and $\beta$ is approximately constant below 50 K, as depicted in Fig. 5(b). We note that the evidence of SC transition was not detected in the temperature dependence of $1/T_1$, while zero resistivity was observed in $\rho$. These experimental facts suggest the suppression of weak superconductivity ($T_c \sim 7$ K) in the heavily overdoped (OD) region under the magnetic fields. 
%The residual superconductivity with the small SC volume fraction could show the sizable Meissner signal. 
%
Moreover, we clarified the linear relation between the $1/T_1$ and $T$ (Korringa law) above 10 K, indicating that the present OA sample is a weakly correlated metal, similar to the OD LSCO. 
Therefore, the magnetic state of LESCO varies from the SG state to the paramagnetic state of the OD region by oxidation annealing. Based on the phase diagram of LSCO, this variation in the ground state is drastic. That is, giving $p$ = 2$\delta$ relation according to the charge neutrality of the sample and $\delta$ of $\sim$ 0.02 for the present LESCO, the expected gain of $p$ owing to the annealing is $\sim$ 0.04. 
This increased amount of $p$ is inadequately small for inducing the variation from the SG phase to the OD phase in LSCO. 
However, a recent oxygen $K$-edge XAS measurement reported the value of $p$ as 0.07 and 0.19 in the AS and OA LESCO with $x$ = 0.26, respectively\cite{Asano_2020}, which was consistent with the drastic variation of magnetism displayed in this study. 
The significant increase in the value of $p$ suggests that the influences of oxidation annealing in $T$$^*$-type compound contained an alternative aspect beyond the hole-doping effect expected from the charge neutrality. %
It is worth mentioning that $T$$^{\prime}$-type LECO exhibited a significant increase of electron density due to the oxygen reduction annealing\cite{Asano_2020arXiv} associated with the alteration from the insulating state to the SC state. 

\color{black}  

%Fig.5%%%%%%%%%%%%%%%%%%%%%%%%%%%%%
\begin{figure}
\begin{center}
\includegraphics[width=8cm,clip]{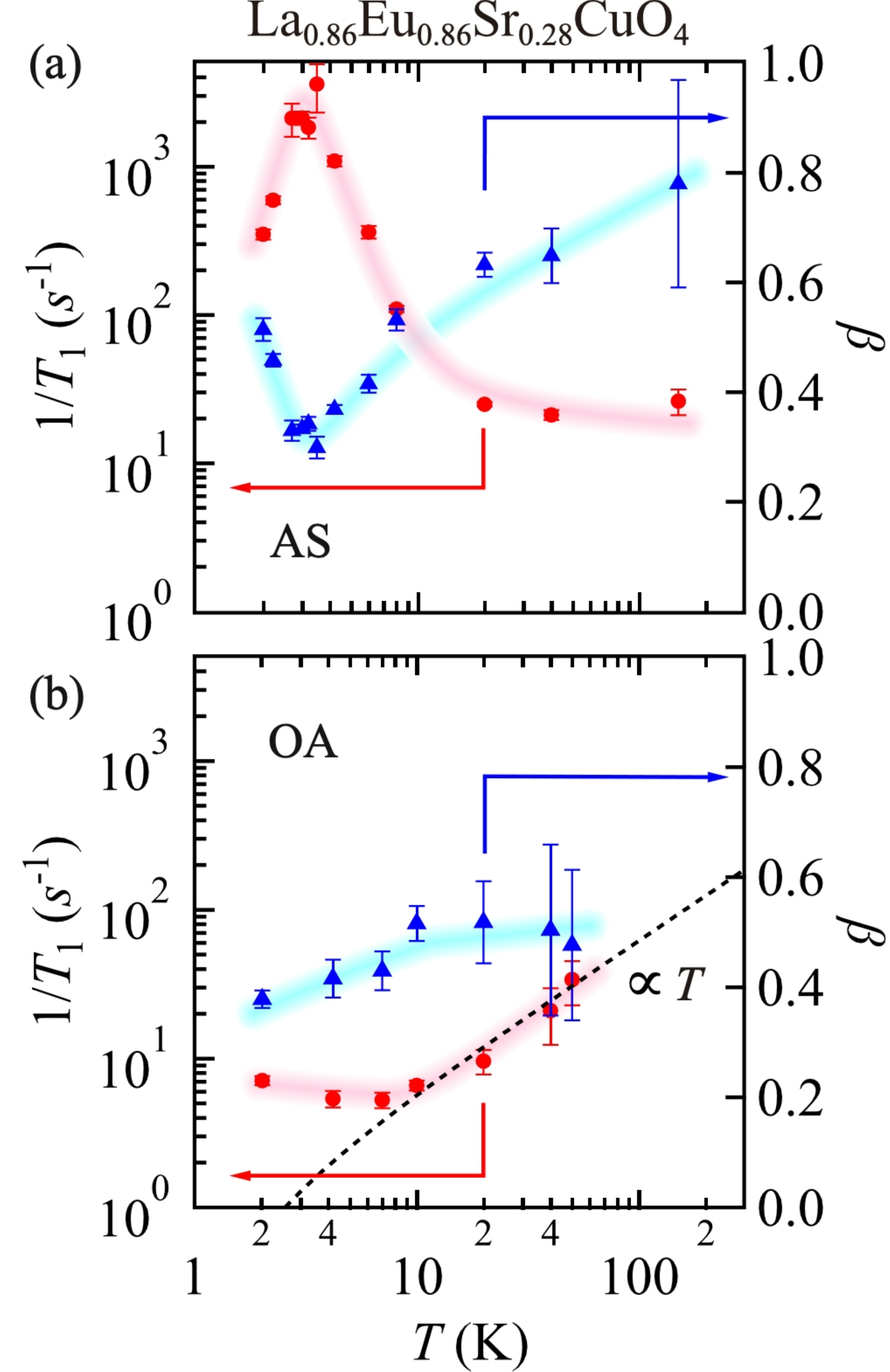}% Here is how to import EPS art
\caption{\label{fig:T1rho} (Color online) Temperature dependence of $1/T_1$ and $\beta$ for (a) AS and (b) OA La$_{0.86}$Eu$_{0.86}$Sr$_{0.28}$CuO$_4$ at 6.22 T. Solid lines are a guide to the eye. 
}
\end{center}
\end{figure}

\subsection{\label{sec:level3}Relationship between spin-dynamics and doped carriers}
Finally, the relationship between the spin-dynamics and the mobility of carrier in AS LESCO is discussed herein. 
Following the prior research on LD LSCO~\cite{Ishida_2004}, we plotted $1/T_1$ and $\rho$ against $1/T$ for AS LESCO. As can be observed from Fig. 6, the $1/T_1$ and $\rho$ displays a similar behavior in the range of $1/T$ between 0.05 ($T$ = 20 K) and 0.33 ($T$ = 3 K). 
Upon applying the relations of $\rho \varpropto \exp (E_{g}/k_{B}T)$ and  $1/T_{1} \varpropto \exp (2J/k_{\rm B}T)$ to the observed results, we estimated $2J/k_{\rm B}$ = 7.8 $\pm$ 0.7 and $E_{\rm g}/k_{\rm B}$ = 6.0 $\pm$ 0.1, respectively.  
Thus, the energy of the transport gap ($E_{g}$) and the characteristic energy of the spin-dynamics (2$J$) in the AS sample are comparable. 
Such a similar behavior in $\rho$ and $1/T_1$ as a function of temperature and the comparable energy scale of two quantities are reported for LD LSCO for $\leq$ 0.024. 
Based on these results, Ishida {\it et al}. claimed that the low-energy spin-dynamics in LD LSCO can be attributed to the mobility of nearly localized carriers, indicating a coupling between the spin and charge degrees of freedom. The spin fluctuations reduced owing to the localization of the holes with the decreasing temperature. 
The inset in Fig. 6 represents the evaluated $E_{g}/k_{B}T$ and $2J/k_{\rm B}$ for the AS LESCO together with those for LSCO as a function of $p$. %
The $E_{g}/k_{B}T$ and $2J/k_{\rm B}$ for the AS LESCO locates almost on the extrapolated position for the LSCO results. 
The consistent results for the two systems suggest that a similar magnetic ground-state to LSCO is realized in the LD region of the AS LESCO, and the origin of the SG-like magnetism can be attributed to the suppression of mobility of doped holes at low temperatures. 

%Fig.6%%%%%%%%%%%%%%%%%%%%%%%%%%%%%
\begin{figure}
\begin{center}
\includegraphics[width=8cm,clip]{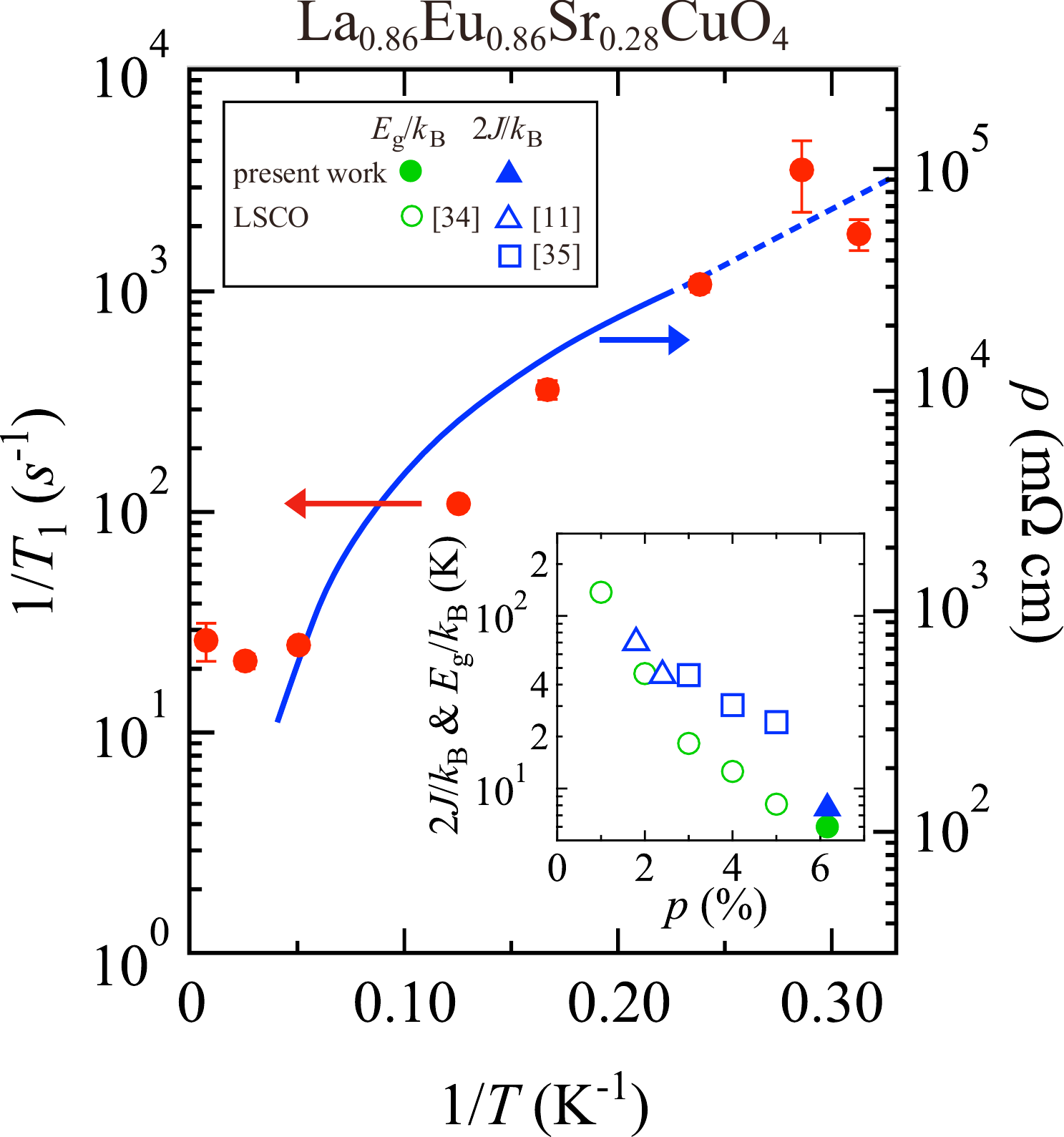}% Here is how to import EPS art
\caption{\label{fig:T1rho} (Color online)  $1/T_1$ (red circles) and $\rho$ (blue line) for AS La$_{0.86}$Eu$_{0.86}$Sr$_{0.28}$CuO$_4$ as a function of $1/T$. Blue dot line indicates extrapolation of $\rho$ at low temperatures. 
Inset represents the hole concentration dependence of $2J/k_{B}$ and $E_{g}/k_{B}$ summarized for La$_{0.86}$Eu$_{0.86}$Sr$_{0.28}$CuO$_4$ and La$_{2-x}$Sr$_{x}$CuO$_4$\cite{Ishida_2004, Ando_2001, Cho_1992}. 
}
\end{center}
\end{figure}

\section{\label{sec:level4}Summary}
In this study, we performed $^{139}$La-NMR measurements on polycrystalline AS and OA $T^*$-type LESCO with $x$ = 0.28. The $^{139}$La-NMR spectrum of LESCO comprises sharp and board components that are consistent with the superposition of the spectra of $T$-type and $T$$^{\prime}$-type cuprates. %Although the SC transition was detected from the resistivity measurements, the absence of the SC anomaly in the NMR measurement suggests that the superconductivity with a small SC volume fraction was present in the OA LESCO under the magnetic fields. 
Moreover, $1/T_1$ in the AS sample increases below 20 K and achieves a maximum at 3 K, implying the appearance of a magnetic order at low temperatures. The logarithmic behavior of $1/T_1$ below 20 K was similar to the semiconducting behavior of resistivity. This similar temperature dependence suggested that the spin-dynamics were attributed to the nearly localized carriers. In the OA sample, no evidence of the magnetic order could be detected, and we confirmed the Korringa relation ($1/T_1 \propto T$) above 10 K. Thus, this NMR study revealed that in $T$*-type LESCO with $x$ = 0.28, the SG state coupled with holes can turn to the weakly correlated metallic state due to the oxidation annealing.

\section*{\label{sec:level4}Acknowledgment}
We thank H. Fukazawa, Y. Ikeda, Y. Kohori, Y. Maeno, Y. Nambu and S. Yonezawa for stimulating discussions, M. Ohkawara for his technical support at HERMES. 
The experiments at JRR-3 was carried out under the general user program managed by the Institute for Solid State Physics, the University of Tokyo (Proposal Nos. 21410 and 21579), and supported by the Center of Neutron Science for Advanced Materials, Institute for Materials Research, Tohoku University. 
This work was financially supported by JSPS/MEXT Grants-in-Aids for Scientific Research (KAKENHI) Grant Nos. JP15H05745, JP16H02125, JP17K14339, JP19K23417, and JP21K13870.  Grant Numbers JP15H05882, JP15H05883, JP15H05884, JP18H04310 (J-Physics).


\begin{thebibliography}{10}

\bibitem{Uchida_1987} 
S. Uchida, H. Takagi, K. Kitazawa, and S. Tanaka, Jpn. J. Appl. Phys. {\bf 26}, L1 (1987).
\bibitem{Takagi_1987} 
H. Takagi, S. Uchida, K. Kitazawa, and S. Tanaka, Jpn. J. Appl. Phys. {\bf 26}, L123 (1987).
\bibitem{Tokura_1989} 
Y. Tokura, H. Takagi, S. Uchida, Nature {\bf 337}, 345 (1989).
\bibitem{Akimitsu_1988}
J.~Akimitsu, S.~Suzuki, M.~Watanabe, and H.~Sawa, Jpn. J. Appl. Phys. {\bfseries 27}, L1859 (1988).
\bibitem{Arima1991} T. Arima, T. Kikuchi, M. Kasuya, S. Koshihara, Y. Tokura, T. Ido, and S. Uchida, Phys. Rev. B {\bf 44}, 917 (1991).
\bibitem{Tshukada2006} A. Tsukada, H. Shibata, M. Noda, H. Yamamoto, M. Naito, Physica C {\bf 445-448}, 94 (2006). 
\bibitem{Ikeda2009} M. Ikeda, T. Yoshida, A. Fujimori, M. Kubota, K. Ono, H. Das, T. Saha-Dasgupta, K. Unozawa, Y. Kaga, T. Sasagawa, and H. Takagi, Phys. Rev. B {\bf 80}, 014510 (2009).

\bibitem{Yamada_1998}
K.~Yamada, C.~H. Lee, K.~Kurahashi, J.~Wada, S.~Wakimoto, S.~Ueki, H.~Kimura, Y.~Endoh, S.~Hosoya, G.~Shirane, R.~J. Birgeneau, M.~Greven, M.~A. Kastner, and Y.~J. Kim, Phys. Rev. B {\bfseries 57}, 6165 (1998).
\bibitem{Fujita_2012}
M.~Fujita, H.~Hiraka, M.~Matsuda, M.~Matsuura, J.~M.~Tranquada, S.~Wakimoto, G.~Xu, and K.~Yamada, J. Phys. Soc. Jpn. {\bfseries 81}, 011007 (2012).
\bibitem{Weidinger_1989}
A. Weidinger, Ch. Niedermayer, A. Golnik, R. Simon, E. Recknagel, J. I. Budnick, B. Chamberland, and C. Baines, Phys. Rev. Lett. 62, 102 (1989).
\bibitem{Julien_2003}
M.-H. Julien, Physica B, {\bf 329-333}, 693 (2003). 
%
%M.-H. Julien, F.~Borsa, P.~Carretta, M.~Horvati\ifmmode~\acute{c}\else\'{c}\fi{}, C.~Berthier, and C.~T. Lin,  Phys. Rev. Lett. {\bfseries 83}, 604 (1999).
\bibitem{Curro_2000}
N.~J. Curro, P.~C. Hammel, B.~J. Suh, M.~H\"ucker, B.~B\"uchner, U.~Ammerahl, and A.~Revcolevschi, Phys. Rev. Lett. {\bfseries 85}, 642 (2000).
\bibitem{Hunt_2001}
A.~W. Hunt, P.~M. Singer, A.~F. Cederstr\"om, and T.~Imai, Phys. Rev. B {\bfseries 64}, 134525 (2001).

\bibitem{Ishida_2004}
K.~Ishida, H.~Aya, Y.~Tokunaga, H.~Kotegawa, Y.~Kitaoka, M.~Fujita, and K.~Yamada, Phys. Rev. Lett. {\bfseries 92}, 257001 (2004).
\bibitem{Hucker_2004}
M. H\"{u}cker, H.-H. Klauss, and B. B\"{u}chner, Phys. Rev. B {\bf 70}, 220507(R) (2004).
\bibitem{Tranquada_1995}
J. M. Tranquada, B. J. Sternlieb, J. D. Axe, Y. Nakamura, and S. Uchida, Nature {\bf 375}, 561 (1995).
\bibitem{Wen2019} J.-J. Wen, H. Huang, S.-J. Lee, H. Jang, J. Knight, Y. S. Lee, M. Fujita, K. M. Suzuki, S. Asano, S. A. Kivelson, C.-C. Kao, and J.-S. Lee, Nature Commun. {\bf 10}, 3269 (2019).
\bibitem{Singer2020} P. M. Singer, A. Arsenault, T. Imai, and M. Fujita, Phys. Rev. B {\bf 101}, 174508 (2020).
\bibitem{Fradkin_2015}
E. Fradkin, S. A. Kivelson, and J. M. Tranquada, Rev. Mod. Phys. {\bf 87}, 457 (2015).
\bibitem{Agterberg_2020}
D. F. Agterberg, J.C. S\'{e}amus Davis, S. D. Edkins, E. Fradkin, D. J. Van Harlingen, S. A. Kivelson, P. A. Lee, L. Radzihovsky, J. M. Tranquada, and Y. Wang, Annu. Rev. Condens. Matter Phys., {\bf 11}, 231 (2020).
\bibitem{Huang_2021}
H. Huang, S.-J. Lee, Y. Ikeda, T. Taniguchi, M. Takahama, C.-C. Kao, M. Fujita, and J.-S. Lee, Phys. Rev. Lett. {\bf 126}, 167001 (2021).

\bibitem{Sawa_1988}
H.~Sawa, H.~Fujiki, K.~Tomimoto, and J.~Akimitsu, Jpn. J. Appl. Phys. {\bfseries 27} L830 (1988).
\bibitem{Akimitsu_1989}
J.~Akimitsu, H.~Sawa, T.~Kobayashi, H.~Fujiki, and Y.~Yamada, J. Phys. Soc. Jpn. {\bfseries 58} 2646 (1989).

\bibitem{Fujita_2018}
M.~Fujita, K.~M. Suzuki, S.~Asano, A.~Koda, H.~Okabe, and R.~Kadono, JPS Conf. Proc. {\bfseries 21}, 011005 (2018).
\bibitem{Asano_2019}
S.~Asano, K.~M. Suzuki, K.~Kudo, I.~Watanabe, A.~Koda, R.~Kadono, T.~Noji, Y.~Koike, T.~Taniguchi, S.~Kitagawa, K.~Ishida, and M.~Fujita, J. Phys. Soc. Jpn. {\bfseries 88}, 084709 (2019).
\bibitem{Asano_2020}
S.~Asano, K.~Ishii, K.~Yamagami, J.~Miyawaki, Y.~Harada, and M.~Fujita, J. Phys. Soc. Jpn. {\bfseries 89}, 075002 (2020).

\bibitem{Matsumoto_2009}
O.~Matsumoto, A.~Utsuki, A.~Tsukada, H.~Yamamoto, T.~Manabe, and M.~Naito, Physica C: Superconductivity {\bfseries 469} 924 (2009).
\bibitem{Takamatsu_2012}
T.~Takamatsu, M.~Kato, T.~Noji, and Y.~Koike, Appl. Phys. Express {\bfseries 5} 073101 (2012).
%\bibitem{Takamatsu_2014}
%T.~Takamatsu, M.~Kato, T.~Noji, and Y.~Koike: Phys. Proc. {\bfseries 58} (2014) 46. 
\bibitem{Sunohara_2020}
T. Sunohara, T. Kawamata, K. Shiosaka, T. Takamatsu, T. Noji, M. Kato, and Y. Koike, J. Phys. Soc. Jpn. {\bf 89}, 014701 (2020).


%\bibitem{Weber_2010_Natphys}
%C.~Weber, K.~Haule, and G.~Kotliar: Nature Phys. {\bfseries 6} (2010) 574.
%\bibitem{Weber_2010_PRB}
%C.~Weber, K.~Haule, and G.~Kotliar: Phys. Rev. B {\bfseries 82} (2010) 125107.

%\bibitem{Arai_1999}
%M.~Arai, T.~Nishijima, Y.~Endoh, T.~Egami, S.~Tajima, K.~Tomimoto, Y.~Shiohara, M.~Takahashi, A.~Garrett, and S.~M. Bennington: Physical Review Letters {\bfseries 83} (1999) 608.

%\bibitem{TranquadaNature2004}
%J.~M. Tranquada, H.~Woo, T.~G. Perring, H.~Goka, G.~D. Gu, G.~Xu, F.~M., and K.~Yamada: Nature {\bfseries 429} (2004) 534.

%\bibitem{HaydenNature2004}
%S.~M. Hayden, H.~A. Mook, P.~Dai, T.~G. Perring, and F.~Dogan: Nature {\bfseries 429} (2004) 531.



\bibitem{Jang_2016}
S.~W. Jang, H.~Sakakibara, H.~Kino, T.~Kotani, K.~Kuroki, and M.~J. Han, Sci. Rep. {\bfseries 6} 33397 (2016).

\bibitem{Adachi_2013}
T.~Adachi, Y.~Mori, A.~Takahashi, M.~Kato, T.~Nishizaki, T.~Sasaki, N.~Kobayashi, and Y.~Koike, J. Phys. Soc. Jpn.  {\bfseries 82} 063713 (2013).

\bibitem{Fukazawa_2017}
H.~Fukazawa, S.~Ishiyama, M.~Goto, S.~Kanamaru, K.~Ohashi, T.~Kawamata, T.~Adachi, M.~Hirata, T.~Sasaki, Y.~Koike, and Y.~Kohori, Physica C, {\bfseries 541} 30 (2017).


\bibitem{Fisk_1989}
Z.~Fisk, S.~W. Cheong, J.~D. Thompson, M.~F. Hundley, R.~B. Schwarz, G.~H. Kwei, and J.~E. Schirbe, Physica C, {\bfseries 162-164} 1681 (1989).

\bibitem{Cheong_1989}
S.~W. Cheong, Z.~Fisk, J.~D. Thompson, and R.~B. Schwarz, Physica C, {\bfseries 159} 407 (1989).


\bibitem{Izumi_1989}
F.~Izumi, E.~Takayama-Muromachi, A.~Fujimori, T.~Kamiyama, H.~Asano, J.~Akimitsu, and H.~Sawa, Physica C {\bfseries 158} 440 (1989).

\bibitem{Bordet_1990}
P.~Bordet, S.~W. Cheong, Z.~Fisk, T.~Fournier, J.~L. Hodeau, M.~Marezio, A.~Santoro, and A.~Varela, Physica C, {\bfseries 171} 468 (1990).

\bibitem{KitaokaJPSJ1987}
Y.~Kitaoka, S.~Hiramatsu, K.~Ishida, T.~Kohara, and K.~Asayama, J. Phys. Soc. Jpn {\bfseries 56} (1987) 3024.

\bibitem{Suter_1998}
A.~Suter, M.~Mali, J.~Roos, and D.~Brinkmann, J. Phys. Cond. Matt. {\bfseries 10} 5977 (1998).

\bibitem{Ando_2001}
Y.~Ando, A.~N.~Lavrov, S.~Komiya, K.~Segawa, and X.~F.~Sun, Phys. Rev. lett. {\bfseries 87} 017001 (2001).

\bibitem{Cho_1992}
J.~H.~Cho, F.~Borsa, D.~C.~Johnston, and D.~R.~Torgeson, Phys. Rev. B. {\bfseries 46} 3179 (1992).

\bibitem{Asano_2020arXiv}
S.~Asano, K.~Ishii, D.~Matsumura, T.~Tsuji, K.~Kubo, T.~Taniguchi, S.~Saito, T.~Sunohara, T.~Kawamata, Y.~Koike, and M.~Fujita, arXiv 2005.10681 (2020).

\end{thebibliography}
\end{document}